\def\o{\over}
\def\Ar{\rightarrow}
\def\bar{\overline}

\def\a{\alpha}

\def\n{\nu}
\def\m{\mu}

\def\e{\epsilon}

\def\th{\theta}

\def\Re{{\rm Re}}
\def\Im{{\rm Im}}

\def\bar{\overline}

\def\G{{\rm GeV}}

\def\eV{{\rm eV}}

\documentstyle [12pt,epsf]{article}
\begin{document}
\baselineskip=24.5pt
\setcounter{page}{1}
\thispagestyle{empty}
\topskip 2.5  cm
%\topskip 0.5  cm
%\begin{flushright}
%\begin{tabular}{c c}
%& {\normalsize hep-ph/   EHU-1}\\
%& March 1998
%\end{tabular}
%\end{flushright}
\vspace{1 cm}
\centerline{\LARGE\bf Indirect Search for $CP$ Violation}
 \centerline{\LARGE\bf in Neutrino Oscillations}

\vskip 1.5 cm
\centerline{{\large \bf Morimitsu TANIMOTO}
  \footnote{E-mail address: tanimoto@edserv.ed.ehime-u.ac.jp}
   }
\vskip 0.8 cm
 \centerline{ \it{Science Education Laboratory, Ehime University, 
 790-8577 Matsuyama, JAPAN}}
\vskip 3 cm
\centerline{\bf ABSTRACT}\par
\vskip 0.5 cm
  We  propose  the indirect search for the $CP$ violating phase
  in the long baseline $\n_\m \Ar\n_e$ oscillation experiment, 
 in which two scenarios of
  the neutrino mass hierarchy are discussed in the three family model.
   The $CP$ violatig phase effect is clearly found 
 in  the scenario:  the LSND data plus the  atmospheric neutrino deficit.
 The phase dependence of the oscillation
 probability is explicitly shown by using  typical  parameters of the
 K2K experiment.
	 The matter effect is negligibly small.
	 In order to select the  scenario of the neutrino mass hierarchy, the
measurement of  the $\n_e\Ar\n_\tau$
oscillation is also proposed  in the long baseline  experiment.
  
\vskip 0.5 cm
 \newpage
%%%%%%%%%%%%%%%%%%%%%%%%%%%%%%%%%%%%%%%%%%%%%%%%%%%%%%%%%%%%%%%%%%%%%%%%%%%%%%%
%%%%%%%%%%%%%%%%%%%%%%%%%%%%%%%%%%%%%%%%%%%%%%%%%%%%%%%%%%%%%%%%%%%%%%%%%%%%%%%
\topskip 0.  cm
\section{Introduction}
 Neutrino flavor oscillations provide 
  information of the fundamental  property of neutrinos  such as  masses, 
flavor mixings and  $CP$ violating phase.  
  In these years, there is growing  experimental evidences of the neutrino
oscillations.
  The exciting one is the atmospheric neutrino deficit \cite{Atm1}$\sim$\cite{Atm3}
    as well as the solar neutrino deficit \cite{solar}.
	   Super-Kamiokande \cite{SKam} also presented the large
neutrino flavor oscillation in  atmospheric neutrinos.
 Furthermore,	a  new stage is represented by the long baseline(LBL) neutrino oscillation experiments \cite{long}.
   The first LBL  reactor experiment CHOOZ has already
reported a  bound  of the neutrino oscillation \cite{CHOOZ},
    which gives a strong constraint of the flavor mixing pattern.
 The LBL accelerator  experiment K2K \cite{K2K} is planned to begin taking data
 in the next year,  whereas the MINOS \cite{MINOS} and ICARUS \cite{ICARUS}
  experiments will start in the first year of the next century.
  Those LBL experiments will clarify  masses,
  flavor mixings and $CP$ violation of neutrinos. 
  
  Some authors \cite{CP1} have already discussed  possibilities of observing $CP$ violation
   in the LBL experiments by measuring 
    the differences of the transition probabilities between $CP$-conjugate
    channels\cite{CP2},  which originates from the phase of the neutrino mixing matrix,
	 such as $\n_\m \Ar\n_\tau$ and $\bar \n_\m \Ar \bar \n_\tau$.
	However, the direct measurement is very difficult in the planned  LBL experiments
 since the  magnitude of its difference is usually expected at most $0.01$ and the difference of
	  energy distributions of neutrino beams $\n_\m$ and
$\bar\n_\m$  disturbs this measurement
	  in the order of  ${\cal O}(0.01)$.
	  
	  In this paper, we propose the indirect search for the $CP$
violating phase
  in the LBL experiments combined with results of the short baseline(SBL) 
   experiments. 
    The SBL experiments  give important constraints for  neutrino mixing
    angles.
	The tentative indication has been already given by the LSND experiment \cite{LSND},
   which is an accelerator experiment for $\n_\m\Ar\n_e(\bar \n_\m\Ar \bar \n_e)$.
   The  CHORUS and NOMAD experiments \cite{CHONOM}
   have  reported the  new bound for $\n_\m\Ar\n_\tau$ oscillation,
   which has already improved the E531 result \cite{E531}.
   The  KARMEN experiment \cite{KARMEN}
 is also searching for the $\n_\m\Ar\n_e(\bar \n_\m\Ar \bar \n_e)$ oscillation 
 as well as LSND. 
 The Bugey \cite{Bugey} and Krasnoyarsk \cite{Kras} reactor experiments and 
	  CDHS \cite{CDHS} and CCFR \cite{CCFR} accelerator experiments have already given
	bounds for the neutrino mixing parameters as well as E776 \cite{E776}.
	
Taking account of those data,  we study the $CP$ violatig phase effect
	 of  the $\n_\m \Ar\n_e$ oscillation in the LBL experiments.
 Bilenky, Giunti and Grimus \cite{BGG} have already discussed the bound of
 this transition probability, which is  
	independent of the $CP$ violatig phase.
	   By investigating the phase effect of the bound,
 we provide a possibility to observe the $CP$ violation in the neutrino oscillation indirectly.

%%%%%%%%%%%%%%%%%%%%%%%%%%%%%%%%%%%%%%%%%%%%%%
 \section{Mass and mixing  patterns and LBL experiments}
 
  Let us start with discussing the recent results of atmospheric neutrinos
	  at Super-Kamiokande \cite{SKam}, which
  suggests $\nu_\m \Ar \nu_\tau$ oscillation with the large mixing.
 Since the CHOOZ result \cite{CHOOZ}
 excludes the large neutrino oscillation of $\nu_\mu \Ar \nu_e$, 
 the large mixing between  $\nu_\m$ and $\nu_\tau$  is a reasonable interpretation
  for the atmospheric $\nu_\mu$ deficit.
  Our starting point as to neutrino mixings is
   the large $\nu_\m \Ar \nu_\tau$ oscillation with 
 \begin{equation}
  \Delta m^2_{\rm atm}\simeq 5\times 10^{-3} \eV^2 \ , \qquad\quad
 \sin^2 2\th_{\rm atm} \simeq 1 \ ,
  \label{atm}
 \end{equation}
\noindent which
  constrain   neutrino oscillations in the LBL experiments.
 The considerable large oscillations are expected  for the $\n_\m \Ar \n_\tau$ process
  in K2K \cite{K2K}, MINOS \cite{MINOS} and ICARUS \cite{ICARUS}
experiments.
 This large  oscillation hardly  depends  on  other mixings and   
the $CP$ violating phase.
 On the other hand, the $\n_\m\Ar \n_e$ oscillation is  small and  depends
  on  the phase and other mixing angles, which are constrained by the SBL experiments and
  the LBL reactor experiments.
  
  In order to give the formulation of the neutrino oscillation
probability for  the LBL $\n_\m\Ar \n_e$ 
   experiment, we should discuss  at first the mass and mixing
pattern of neutrinos.
 Other  possible evidences in favor of neutrino oscillations are
 the solar neutrino and the accelerator experiment LSND.
 Three different scales of mass-squared differences 
 cannot be  reconciled with the three family model of neutrinos unless
one of data is disregarded.
 Therefore, one should two possible scenario\cite{Smi}:
(1) "{\bf sacrifice solar neutrino}" and (2) "{\bf sacrifice LSND}".
 By stretching the data, the other scenario  is still available
    as discussed by Cardall and Fuller \cite{CaFu}.
 This scenario appears to emerge naturally as the most likely solution to all
 oscillation evidences, however,  contradicts  with the zenith angle
 dependence of multi-GeV atmospheric data in the Kamiokande and
 Super-Kamiokande experiments \cite{SKam}.
 So, we do not discuss this scenario in our  paper. 
  If one would like to sacrifice no data, the sterile neutrino  should be
   included in the analyses, which is out of scope in our paper.

%%%%%%%%%%%%%%%%%%%%%%%%%%%%%%%%%%%%%%%%%%%%%%%%%%%%%%%%%%%
  In the scenario  (1) {\bf "sacrifice solar neutrino"},
   we take $\Delta m_{31}^2\simeq\Delta m_{21}^2\simeq \Delta m^2_{\rm LSND}$ and
$\Delta m_{32}^2 \simeq \Delta m^2_{\rm atm}$($m_3\simeq m_2\gg m_1$).  
 The natural  pattern of the neutrino mixing matrix $U$ \cite{Masa} is
 \begin{equation} 
 U \simeq \left (\matrix{1 & \e_1 & \e_2 \cr \e_3 & c & s
\cr\e_4 & -s & c \cr} \right ) \ ,
\label{U1}
\end{equation}
where $c\equiv\cos\th\simeq 1/\sqrt{2}$ and $s\equiv \sin\th\simeq
1/\sqrt{2}$ are fixed  by taking account  of eq.(\ref{atm}),
and small values $\e_i$'s are constrained from the SBL  experiments.
 We do not consider the mass hierarchy $m_3\gg m_2\gg m_1$, which is unfavourable  in constructing the neutrino mass matrix without fine-tuning 
 in the case of  the large mixing \cite{Masa}.

%%%%%%%%%%%%%%%%%%%%%%%%%%%%%%%%%%%%%%%%%%%%%%%%%%%%%%%
In the scenario  (2) {\bf "sacrifice LSND"},
  we take
$\Delta m_{32}^2\simeq\Delta m_{31}^2\simeq \Delta m^2_{\rm
atm}$ and     $\Delta m_{21}^2   \simeq \Delta m^2_\odot$(solar neutrino).
 For the MSW small angle solution \cite{MSW}, the mixing pattern is
the same one in eq.(\ref{U1}).
Another pattern  is equivalent to an exchange of the first and second
columns in the mixing matrix. 
In the case of the just-so solution and the MSW
large angle solution, the mixing matrix pattern
 is somewhat complicated.  Taking account of the  data of the CHOOZ experiment,
 Bilenky and Giunti \cite{Giunti} have obtained a typical  mixing
pattern, on which we will comment later.

%%%%%%%%%%%%%%%%%%%%%%%%%%%%%%%%%%%%%%
 Now, in terms of the standard parametrization of the mixing matrix $U$ \cite{PDG},
 \begin{equation}  
 U= \left (\matrix{ c_{13} c_{12} & c_{13} s_{12} &  s_{13} e^{-i \phi}\cr 
  -c_{23}s_{12}-s_{23}s_{13}c_{12}e^{i \phi} & c_{23}c_{12}-s_{23}s_{13}s_{12}e^{i \phi} & 
                       s_{23}c_{13} \cr
  s_{23}s_{12}-c_{23}s_{13}c_{12}e^{i \phi} & -s_{23}c_{12}-c_{23}s_{13}s_{12}e^{i \phi} & 
                       c_{23}c_{13} \cr} \right ) \ ,
\end{equation} 
\noindent   where  $s_{ij}\equiv \sin{\theta_{ij}}$ and $c_{ij}\equiv \cos{\theta_{ij}}$
 are vacuum mixings, and  $\phi$ is the $CP$ violating phase,
 the neutrino oscillation probability for the LBL $\n_\m\Ar \n_e$
experiment is given  by
   \begin{equation}        
 P(\n_\m\Ar \n_e) \simeq  2|U_{e 1}|^2 |U_{\m 1}|^2-4 c_{23}s_{23}s_{12}s_{13}\cos{\phi}
            \sin^2{\Delta m^2_{32} L \o 4 E} 
			 - 2 c_{23}s_{23}s_{12}s_{13}\sin{\phi} \  S_{CP}
	\label{P1}
\end{equation}
\noindent with
   \begin{equation}        
  |U_{e 1}|^2 |U_{\m 1}|^2 \simeq
   c_{23}^2 s_{12}^2 + s_{23}^2 s_{13}^2 + 2 c_{23}s_{23}s_{12}
s_{13}\cos{\phi} \ ,
   \label{P11}
\end{equation}
 for the scenario (1),
          and 
   \begin{equation}        
 P(\n_\m\Ar \n_e) \simeq  4 s_{23}^2 s_{13}^2 \sin^2{\Delta m^2_{31} L \o 4 E} 
			 - 2 c_{23}s_{23}s_{12}s_{13}\sin{\phi} \ S_{CP}
	\label{P2}
   \end{equation}
\noindent  for the scenario (2),
where
  \begin{equation} 
   S_{CP}=\sin {\Delta m^2_{12} L \o 2 E}+ \sin {\Delta m^2_{23} L \o 2 E}+
          \sin {\Delta m^2_{31} L \o 2 E} \ .
 \end{equation} 
  The oscillatory term  $S_{CP}$ is approximately reduced to  
$\sin ({\Delta m^2_{23} L/2 E})$
   for the scenario (1), because  
   $\sin ({\Delta m^2_{12} L/2 E})$ and $\sin ({\Delta m^2_{31} L/2
E})$
 can be  replaced by the average value $0$ in the  LBL experiments.
	For the scenario (2),  $S_{CP}$ is suppressed
  such as $S_{CP}\sim 0.01$,
	 so the second term  in  eq.(\ref{P2}) can be  neglected.
	
	Thus, the probability in eq.(\ref{P1}) strongly depends on the $CP$ violating phase,
	 while the one in eq.(\ref{P2}) hardly  depends on it. 
	 The $CP$ violating phase effect could be found  in the LBL
$\n_\m\Ar \n_e$ experiment  if the scenario (1) is true. On the other hand, if the scenario (2) is true, it is impossible to
	 determine the  $CP$ violating phase by $\n_\m\Ar \n_e$.

 %%%%%%%%%%%%%%
 %%%%%%%%%%%%%%%%%%%%%%%%%%%%%%%%%%%%%%%%%%%%%%%%%%%%%%%%%%
 \section{Constraints from reactor and accelerator experiments} 
 
 In our analyses,  $s_{23}\simeq 1/\sqrt{2}$ is fixed, but   
  $s_{12}$, $s_{13}$ and $\phi$  still remain  as  free patrameters.
  Those mixing parematers are constrained from other experiments.
 %%%%%%%%%%%%%%%%%%%%%%%%%%%%%%%%%%%%%%%%%%%%%%%%%%%%%%%  
 %% Constraints in  scenario (1)
 %%%%%%%%%%%%%%%%%%%%%%%%%%%%%%%%%%%%%%%%%%%%%%%%%%%%%%
 For the scenario (1), the SBL experiments give strong constraints because 
  the distance $L$ corresponds to the neutrino mass scale ${\cal O}(1\eV)$.
  Let us consider   constraints from  disappearance experiments.
 The Bugey \cite{Bugey} and  Krasnoyarsk \cite{Kras} reactor experiments  and 
	  CDHS \cite{CDHS} and CCFR \cite{CCFR} accelerator experiments give bounds for the neutrino
 mixing parameters at the fixed value of $\Delta m^2_{31}$ \cite{Bil}.
In the mixing pattern of eq.(\ref{U1}), the constraint for $U_{e 1}$
is given by using
\begin{equation}        
P(\bar\n_e\Ar \bar\n_e) \simeq  1 - 4 |U_{e 1}|^2 (1 - |U_{e 1}|^2) 
            \sin^2 {\Delta m^2_{31} L \o 4 E} \ ,
\end{equation}
\noindent
where  $\Delta m^2_{31}\gg \Delta m^2_{32}$ is used.
Thus, one obtains  the bound $|U_{e1}|^2\geq a_e$ due to
non-observation of  neutrino oscillations.  For example, we get
   $a_e=0.984$ for $\Delta m^2_{31}=2 \eV^2$.
 This condition is expressed in terms of mixing parameters as follows: 
 
\begin{equation}          
 s_{12}^2+s_{13}^2 -s_{12}^2 s_{13}^2 \leq  B_e \equiv 1-a_e \ .
 \label{cond1}
  \end{equation} 
  
 Constraints from  the  appearance experiments are given by LSND, E776 and CHORUS/NOMAD.
    If the excess of the electron events at LSND are due to the neutrino oscillation, 
the $\bar \n_\m\Ar \bar \n_e$ oscillation
 transition probability is equal to
 $0.31\pm 0.09\pm 0.05\%$ \cite{LSND}. 
  The plot of the LSND favored region of  $\Delta m^2$ vs $\sin^2 2\theta$  allows
    the mass region to be  larger  than $\Delta m^2\simeq 0.04\ \eV^2$.
	However,  Bugey\cite{Bugey} has  already excluded
	 a mass region lower than $\Delta m^2=0.27\ \eV^2$.  On the other hand,	the constraint by  E776 \cite{E776}  has excluded a  region
	larger than $\Delta m^2=2.3\ \eV^2$.
 Thus, the mass squared difference $\Delta m_{31}^2$ is obtained in the range \cite{OYB}:
  \begin{equation}        
     \Delta m_{31}^2\simeq 0.27 \ \eV^2 \sim 2.3 \ \eV^2  \ .
   \end{equation}
   The recent CHORUS experiment for $\n_\m\Ar\n_\tau$  has given the bound  $\sin^2 2\th_{\rm
CHORUS}\leq 1.3\times 10^{-3} \cite{CHONOM}$.
   Those data lead to   constraints of  mixing angles
   by using following equations:
 \begin{eqnarray}   
     P(\n_\m\Ar \n_e) \simeq  4 |U_{e 1}|^2 |U_{\m 1}|^2 
            \sin^2 {\Delta m^2_{31} L \o 4 E} \ ,  \nonumber \\
     P(\n_\m\Ar \n_\tau) \simeq 4 |U_{\m 1}|^2 |U_{\tau 1}|^2 
            \sin^2 {\Delta m^2_{31} L \o 4 E} \ .
 \end{eqnarray}
  Using eq.(\ref{P11}), we get the condition for the LSND data 
  \begin{equation}        
   c_{23}^2 s_{12}^2 + s_{23}^2 s_{13}^2 + 2 c_{23}s_{23}s_{12} s_{13}\cos{\phi}=L_e\equiv 
   {1\o 4}\sin^2 2\th_{\rm LSND}\ ,
   \label{cond2}
\end{equation}
\noindent
where $L_e$ is given  experimentally  for fixed  $\Delta m^2_{31}$.

For the CHORUS bound of $\n_\m\Ar \n_\tau$, we  also obtain a condition
 \begin{equation}        
   s_{23}^2 s_{12}^2 + c_{23}^2 s_{13}^2 - 2 c_{23}s_{23}s_{12} s_{13}\cos{\phi}\leq C_\tau\equiv 
   {1\o 4}\sin^2 2\th_{\rm CHORUS}\ ,
\end{equation}
  \noindent
  which is a weak bound at present since $\sin^2 2\th_{\rm CHORUS}\geq 0.1$ in the case of 
  $\Delta m^2_{31}\leq 2\eV^2$. Actually,  mixing angles  which satisfy 
  eqs.(\ref{cond1}) and (\ref{cond2}) always  satisfy this bound.
  
  The first LBL reactor experiment CHOOZ \cite{CHOOZ} also constrains
 mixing angles
  by the following formula:
   \begin{equation}        
  P(\bar\n_e\Ar \bar\n_e) \simeq 1 - 2 |U_{e 1}|^2 (1 - |U_{e 1}|^2) -
         4 |U_{e 2}|^2 |U_{e 3}|^2 \sin^2 {\Delta m^2_{32} L \o 4 E} \ .
  \end{equation}
  \noindent
  Taking account of $s_{12}\ll 1$ and  $s_{13}\ll 1$, the probability reduces to
   \begin{equation}        
  P(\bar\n_e\Ar \bar\n_e) \simeq 1 - 2 (s_{12}^2 + s_{13}^2) \ .
  \end{equation}
  \noindent
  Since CHOOZ reported  $\sin^2 2\th_{\rm CHOOZ}\leq 0.18$ for large $\Delta m^2$,
   we get a constraint $s_{12}^2 + s_{13}^2\leq 0.18/4$, which is a loose bound
    compared with the one in eq.(\ref{cond1}).

 %%%%%%%%%%%%%%%%%%%%%%%%%%%%%%%%%%%%%%%%%%%%%%%%%%%%%%%  
 %% Constraints in  scenario (2)
 %%%%%%%%%%%%%%%%%%%%%%%%%%%%%%%%%%%%%%%%%%%%%%%%%%%%%%
  Next we consider the scenario (2), in which the SBL experiments do not constrain  mixing 
  angles because the highest mass difference scale is ${\cal O}(0.01\eV^2)$.
  The constraint follows only from  the CHOOZ experiment \cite{CHOOZ},
in which the probability is expressed as
   \begin{equation}        
  P(\bar\n_e\Ar \bar\n_e) \simeq 1 - 
        4 |U_{e 3}|^2 (1-|U_{e 3}|^2) \sin^2 {\Delta m^2_{31} L \o 4 E}
	    \simeq 1 - 4 s_{13}^2 (1 - s_{13}^2) \sin^2 {\Delta m^2_{31} L \o 4 E} \ .
  \end{equation}
  \noindent
  By using the CHOOZ result  $\sin^2 2\th_{\rm CHOOZ}\leq 0.12$ for 
  $\Delta m^2\simeq 5\times 10^{-3} \eV^2$, we get only a constraint $s_{13}^2\leq 0.03$.
 The MSW small angle solution of the solar neutrino \cite{MSW} gives another constraint:
 $s_{12}= 0.02\sim 0.06$.

%%%%%%%%%%%%%%%%%%%%%%%%%%%%%%%%%%%%%%%%%%%%%%%%%%%%%%
 \section{Phase dependence of $P(\n_\m\Ar \n_e)$}   
 
%%%%%%%%%%%%%%%%%%%%%%%%%%%%%%
%%% scenario (1) in P(\n_\m\Ar \n_e)
%%%%%%%%%%%%%%%%%%%%%%%%%%%%%%
 In  the scenario (1),
  eq.(\ref{P1}) is written as 
    \begin{equation}        
 P(\n_\m\Ar \n_e) \simeq  2|U_{e 1}|^2 |U_{\m 1}|^2 - 4 c_{23}s_{23}s_{12}s_{13}
   \sin\left ( {\Delta m^2_{32} L \o 4 E}\right )\sin\left ({\Delta m^2_{32} L \o 4 E} -\phi\right ) \ ,
 \label{Pme}
 \end{equation}
 \noindent
 where the first term in the right hand side is fixed by the LSND data.
 It is remarked that $c_{23}s_{23}s_{12}s_{13}$ determines this oscillation probability
    if $\phi$ is fixed. So, the maximal value of $c_{23}s_{23}s_{12}s_{13}$
 gives the maximal or minimal $P(\n_\m\Ar \n_e)$, which  depends on $\phi$.

	We show how to extract the maximal value of $s_{12}s_{13}$
	 from eqs.(\ref{cond1}) and (\ref{cond2}), taking 
	   $s_{23}=1/\sqrt{2}$  for simplicity. Those conditions turn to 
  \begin{eqnarray}          
      &&s_{12}^2+s_{13}^2 -s_{12}^2 s_{13}^2 \leq  B_e  \ ,  
           \label{Scond1} \\
      &&s_{12}^2 + s_{13}^2 + 2 s_{12} s_{13}\cos{\phi} = 2 L_e \ .
           \label{Scond2}
 \end{eqnarray}
	  The maximal value of $s_{12}s_{13}$ under the condition in eq.(\ref{Scond2})
	 is  obtained by an easy algebraic manipulation  as follows:
  \begin{equation}        
    s_{12}s_{13}\leq {1\o 1+\cos{\phi}} L_e \ .
	\label{sol1}
  \end{equation}
  \noindent
  On the other hand, we get an equation as to $s_{12} s_{13}$ by a subtraction
  of eq.(\ref{Scond1}) and  eq.(\ref{Scond2}):
  \begin{equation}          
     s_{12}^2 s_{13}^2 + 2 s_{12} s_{13}\cos{\phi} + (B_e-2 L_e) \geq 0 \ .
   \label{Scond3}
 \end{equation}
 Taking a positive solution of $s_{12} s_{13}$, we get
   \begin{equation}          
    s_{12}s_{13}\leq -\cos{\phi} - \sqrt{\cos^2 \phi - (B_e -2 L_e)}  \ .
   \label{sol2}
 \end{equation}
 \noindent
 The true maximum of $s_{12} s_{13}$ should satisfy both conditions 
 of eqs.(\ref{sol1}) and (\ref{sol2}). In conclusion,  we obtain the
 maximal $s_{12} s_{13}$  as follows:
	 \begin{eqnarray}          
        s_{12}s_{13}\leq {1\o 1+\cos{\phi}} L_e  \ \ {\rm for} \ \
	                   \cos\phi \geq -1+{L_e\o B_e} (1+\sqrt{1-B_e}) \ ,
           \nonumber \\
        s_{12}s_{13}\leq -\cos{\phi} - \sqrt{\cos^2 \phi - (B_e -2 L_e)} \ \ {\rm for} \ \
	   \cos\phi \leq -1+{L_e\o B_e} (1+\sqrt{1-B_e}) \ . 
           \label{SOL}
     \end{eqnarray}    
 
 \noindent
 By using this maximal value, we can  predict the allowed region of the 
  oscillation probability of eq.(\ref{Pme}) for the fixed value of  $\phi$.
  We show the allowed region of $P(\n_\m\Ar \n_e)$ versus $\phi$ in fig.1, where 
    $\Delta m^2_{32}=5\times 10^{-3}\eV^2$ and typical  parameters of
the  K2K experiment 
	$L=250{\rm Km}$, $E=1.3\G$ \cite{K2K} are taken.
  We also take $\Delta m^2_{31}=2\eV^2$, which leads to $B_e=0.0183$ and
  $L_e=6.5\times 10^{-4}$. The allowed region is shown  between the
solid curve and the  horizontal dashed line.
   It is noticed that this result is consistent with the one  in 
ref.\cite{BGG}.
   Since the condition of  eq.(\ref{SOL}) is  a general one, it is useful even if the LSND data
   is replaced by  the KARMEN \cite{KARMEN} data in the future.
   
   As seen in fig.1, the probability increases steeply in the region $\phi\geq 100^{\circ}$ and 
   to the maximal value around $0.02$.
   Thus, the observation of $P(\n_\m\Ar \n_e)$ larger than $0.005$ indicates
   the large $CP$ violating phase although
  the $CP$ conserved  limit $\phi=180^\circ$   is still allowed.
  In the case of $\phi=180^\circ$, the observed $P(\n_\m\Ar \n_e)$
	fixes both $s_{12}$ and $s_{13}$.  Then, one can predict the $CP$ conjugated
	 process $\bar\n_\m\Ar \bar\n_e$ for arbitrary neutrino beam energies. 
Thus, the measurement of the  absolute value of $P(\bar\n_\m\Ar \bar\n_e)$ 
	 is a crucial test of the $CP$ violation.
	 In fig.1, we  have also  shown the oscillation probability of  
    $\bar\n_\m\Ar \bar\n_e$ by taking same values of parameters $L$ and $E$.

%%%%%%%%%%%%%%%%%%%%%%%%%%%%%%
%%% scenario (2) in P(\n_\m\Ar \n_e)
%%%%%%%%%%%%%%%%%%%%%%%%%%%%%%
In the scenario (2), substituting a bound $s_{13}^2\leq 0.03$ into eq.(\ref{P2}), we obtain
   \begin{equation}        
      P(\n_\m\Ar \n_e) \leq 0.06 \ ,
   \end{equation}
 \noindent
 which is much larger than the maximal value $0.02$ in the scenario (1).
 In this case, the $CP$ phase dependence  is not found.
 
 We have taken the MSW small angle solution of the solar neutrino 
in the scenario (2).
 In the case of the just-so solution and the MSW
large angle solution, the mixing matrix pattern
 is somewhat complicated.  However, the mixing matrix
 given in ref.\cite{Giunti} hardly  changes our result with  the MSW small
angle solution.

 The observation of $0.02\leq P(\n_\m\Ar \n_e)\leq 0.06$ means that the scenario (2)
  is true,  that is to say, the mass difference scale $\Delta m^2={\cal O}(1\eV^2)$
   is rejected.  This is the same issue in ref.\cite{Shimada}.
 What can one say if its observed probability is lower than $ 0.02$?
The measurement of the $\bar\n_\m\Ar \bar\n_e$ oscillation can select
a  scenario because the $CP$ violating phase effect is found in the  scenario (1). 
 An alternative method is to measure the $\n_e\Ar\n_\tau$
oscillation in the LBL experiment.  For  the scenario (1), the
probability
is different from the  $\n_\m\Ar\n_e$ oscillation as follows:
    \begin{equation}        
 P(\n_e\Ar \n_\tau) \simeq  2|U_{e 1}|^2 |U_{\tau 1}|^2 +4 c_{23}s_{23}s_{12}s_{13}
   \sin\left ( {\Delta m^2_{32} L \o 4 E}\right )\sin\left ({\Delta m^2_{32} L \o 4 E} +\phi\right ) \ ,
 \label{Petau}
 \end{equation}
 \noindent
 while  the  probability is the same one for  the scenario (2)
in the case of  $s_{23}=1/\sqrt{2}$.
Thus, the  $\n_e\Ar\n_\tau$ oscillation can select a  scenario of the
mass hierarchy.

It may be useful to remark that  the observation of $P(\n_\m\Ar \n_e)$ larger than $0.06$ indicates new physics.

%%%%%%%%%%%%%%%%%%%%%%%%%%%%%%%%%%%%%%%%%%%%%%%
%%%%%%%%%%%%%%%%%%%%%%%%%%%%%%%%%%%%%%%%%%%%%%%
\section{Matter effect} 

  Although the distance travelled by neutrinos is less than $1000 {\rm~Km}$ in
 the LBL experiments, those data  include  the background
 matter effect which is not $CP$ invariant \cite{matter}.  
In particular, 
 the matter effect should be carefully analyzed  for  the scenario  (1)
 since the numerical results strongly depend on the $CP$ violating phase.

  \noindent
   Hamiltonian in matter $H(x)$  in weak basis is 
\begin{equation}  
 H = U \left (\matrix{ m_1^2/2E & 0 & 0 \cr 
            0 & m_2^2/2E & 0 \cr
            0 & 0 & m_3^2/2E \cr} \right )U^\dagger +
      \left (\matrix{ a & 0 & 0 \cr 
            0 & 0 & 0 \cr
            0 & 0 & 0 \cr} \right )  \ , 
\end{equation} 
\noindent where
\begin{equation} 
 a=\sqrt{2} G_F n_e \ ,  
 \end{equation} 
 \noindent with
 a constant matter density  $\rho=2.34\ {\rm g/cm}^3$.
 For antineutrinos, effective Hamiltonian is given by
  replacing $a\Ar -a$ and $U \Ar U^*$.
  
  Minakata and Nunokawa have shown the perturbative treatment of the matter effect  
  \cite{CP1}, which is reliable under the mass scale  we consider. 
  In the scenario (1)  with 
$ m_3\simeq m_2 \gg m_1$, the transition probability is
 given:
\begin{eqnarray}
  P(\nu_\m\Ar\nu_e)&=& 2 |V_{e1}(a)|^2|V_{\m1}(a)|^2 
             -4\Re[V_{e2}(a)V^*_{e3}(a)V^*_{\m2}(a)V_{\m3}(a)]   \sin^2{1\o 2}I_{23}   
   \nonumber \\ 
  &+&2 \Im[V_{e2}(a)V^*_{e3}(a)V^*_{\m2}(a)V_{\m3}(a)] \sin I_{23} \ ,
\label{PT}
\end{eqnarray}
\noindent  where 
\begin{equation} 
   V_{\a i}(a)= U_{\a i}-\sum_{j\not=i}^{}2aE{U_{\a j}U^*_{ej}U_{ei}\o  m_j^2-m_i^2} \ ,
   \quad
   I_{ij}=-{\Delta m^2_{ij} L \o 2 E}+(|U_{ei}|^2 - |U_{ej}|^2)aL \ .
   \end{equation} 
   \noindent
   The oscillation probability turns to  a complicated equation, but
    it is easy to estimate the matter oscillation effect by using
    the parameters $E={\cal O}(1\G)$, $a={\cal O}(10^{-13}\eV)$ and  $L=250 {\rm Km}$.
	We have found the leading matter corrections for the vacuum oscillation to be dominated by
	 terms of $2aE/\Delta m^2_{31}$,  $2aEs_{12}^2/\Delta m^2_{32}$,  
	$2aEs_{13}^2/\Delta m^2_{32}$ and  $aL(s_{12}^2-s_{13}^2)$.
	Those corrections are  at most ${\cal O}(10^{-3})$ for the vacuum oscillation probability
	 of $\nu_\m\Ar\nu_e$ in the case of $\Delta m^2_{31}\simeq1 \eV^2$.
	 This result is consistent with the previous numerical one \cite{CP1}.
	 Thus, the matter effect is negligible small.

\section{Conclusions}\par
 
  We have proposed the indirect search for the $CP$ violating phase
  in the LBL experiments combined with results of the SBL experiments in two scenarios of
  the mass hierarchy.
   Taking account of those data,  we have investigated the $CP$ violatig phase effect
	 of  the $\n_\m \Ar\n_e$ oscillation in the LBL experiments for the scenario (1):
	 the LSND data plus the  atmospheric neutrino deficit.
	 For the scenario (2): the atmospheric neutrino deficit plus the solar neutrino deficit,
	 the phase dependence is negligibly small.
	 As an example, we have shown the phase dependence of the oscillation
	 probability by using typical  parameters of the K2K experiment for the scenario (1).
	 The matter effect has been found to be negligibly small.
	  This phase effect on the probability 
   provides a possible method to observe the $CP$ violation in the neutrino oscillation indirectly.   In order to select the  scenario (1) or (2), the
measurement of  the $\n_e\Ar\n_\tau$
oscillation is useful  in the LBL experiment 
as well as  $\bar\n_\m\Ar\bar\n_e$. 
   Even if  our predicted magnitudes are  out of sensitivity in the K2K experiment,
    we can expect the experiments of MINOS and ICARUS.

\newpage
%%%%%%%%%%%%%%%%%%%%%%%%%%%%%%%%%%%%%%%%%%%%%%%%%%%%%%%%%%%%%%%%%%%%%%%%%%%%%%%%%%%%%%%%

 \newpage

%%%%%%%%%%%%%%%  Figure 1  %%%%%%%%%%%%%%%%%
\begin{figure}
\epsfxsize=14 cm
\centerline{\epsfbox{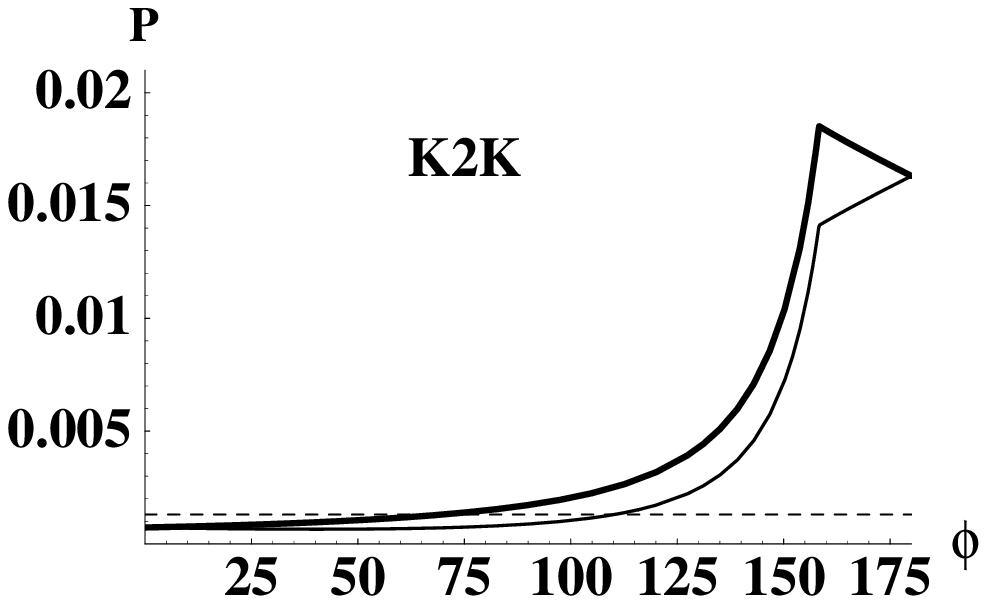}}
\caption{}
	 \end{figure}
Fig. 1: \ \ Predicted bound of the  $\n_\m \Ar\n_e$ oscillation 
in the K2K experiment for the scenario (1), which is denoted
 by the thick solid curve.  
	The $CP$ conjugated  $\bar\n_\m \Ar\bar\n_e$ one 
 is shown by the thin solid curve. 
	The horizontal dashed line denotes $2|U_{e1}|^2|U_{\m1}|^2$,
which is fixed  by the LSND data. Here,
	$\Delta m^2_{31}=2 \eV^2$ and  $\Delta m^2_{32}=5\times 10^{-3} \eV^2$ are taken.
%%%%%%%%%%%%%%%%%%%%%%%%%%%%%%%%%%%%%%%%%%%%% 
\end{document}